# Surrogate–Newton coupling enables fast and reliable steady CFD for out-of-distribution geometries

Mingcheng Lei[1], Weishao Tang[1], Haixin Chen[1], and Yufei Zhang[1,2*]

[1] School of Aerospace Engineering, Tsinghua University, Beijing 100084, China

[2] State Key Laboratory of Advanced Space Propulsion, Tsinghua University, Beijing 100084, China

* Corresponding author, e-mail: zhangyufei@tsinghua.edu.cn

## Abstract

Neural surrogates promise faster physical simulations; however, unreliable predictions for out-of-distribution (OOD) inputs limit their practical applicability. In this study, we develop a surrogate–Newton framework for steady computational fluid dynamics (CFD) that combines learned full-field initialization with residual-controlled Newton–Krylov correction. Convergence analysis and controlled experiments demonstrate the effectiveness of surrogate predictions as initial states for Newton–Krylov solution. On 16,399 transonic flow cases, the framework reduces the median field mean-squared error by more than six orders of magnitude compared with standalone surrogate predictions. In 200 matched timing queries, it reduces the median solver time at termination by a factor of 5.8 relative to uniform-initialized CFD. The framework also extends to three-dimensional wing flows on meshes containing approximately 3.5 million cells, achieving a 5.2-fold median paired speedup at the same target residual ratio. When applied to aerodynamic design optimization, the framework reduces transonic optimization time by a factor of 17.3 compared with CFD, using Top-4 candidate screening. It also yields independently verified airfoil designs for Mars unmanned aerial vehicles under extreme OOD conditions. These results establish surrogate–Newton coupling as a route to reliable and efficient steady CFD for OOD geometries.

## Introduction

Neural surrogates offer a means of accelerating simulations governed by partial differential equations (PDEs)[1] by learning reusable mappings from problem parameters to solution fields.[2–4] In computational fluid dynamics (CFD), this capability could reduce the computational cost of repeated flow evaluations in applications such as aircraft and automotive design,[5] thereby shortening development cycles and enabling broader exploration of candidate configurations.[6,7] However, practical design exploration may extend beyond the geometric variations represented in the training data. Reduced accuracy on such out-of-distribution (OOD) queries can compromise performance assessments and misdirect design decisions, posing a barrier to industrial deployment.[8–10] Realizing these efficiency gains in engineering design therefore requires reliable performance assessment for OOD geometries.

Meeting this requirement is challenging for standalone data-driven models, which rapidly capture global flow structures[11–13] but rely on learned mappings rather than residual-controlled convergence for each new configuration. Their accuracy depends strongly on training coverage, which is constrained by the difficulty of sampling high-dimensional geometry spaces[14,15] and the cost of generating reference flow solutions.[9,16] Physics-informed training incorporates equation residuals and physical constraints into the learning objective to reduce data dependence.[17–19] Although such constraints can improve accuracy and physical consistency, training-time regularization does not by itself guarantee reliable OOD predictions at inference.[18,20]

A complementary strategy applies physical constraints directly at inference,[18,21] refining a surrogate prediction through gradient-based residual minimization[18] or pseudo-time marching in a coupled CFD solver.[22,23] However, using a surrogate prediction as an initial field does not automatically yield a computational speedup: first-order or pseudo-time refinement can still converge slowly on stiff, strongly nonlinear problems such as transonic aerodynamics.[24,25] Recent hybrid frameworks couple neural operators with numerical discretizations[26,27] or warm-start iterative and Newton solvers,[28,29] yet these efforts have largely focused on elliptic PDEs, canonical benchmarks, or solid-mechanics problems. Whether direct full-field surrogate predictions can reliably accelerate coupled Newton solvers in steady compressible turbulent CFD, particularly under strong OOD geometric shift and in realistic engineering problems, remains largely unexplored.

In this study, we develop a surrogate–Newton framework for steady aerodynamic CFD that combines rapid full-field prediction with a fully coupled Newton–Krylov (NK) solution of the discrete steady equations. The surrogate can rapidly capture useful global flow structures and supply an informed initial state under geometric distribution shift, even when its prediction does not meet the accuracy requirements of a final CFD solution. NK offers rapid residual convergence near a steady solution,[24,30] but its effectiveness in stiff nonlinear systems depends strongly on the initial state, and unsuitable initial states can lead to stagnation or divergence.[31,32] The coupling thus uses learned flow structures to help satisfy NK's initialization requirements, whereas NK provides residual-controlled convergence toward a high-accuracy solution. This complementarity offers a route to reliable and efficient steady CFD under geometric distribution shift.

We assess the reliability, accuracy, and computational efficiency of the framework under geometric distribution shift using a large-scale transonic airfoil benchmark comprising geometries sampled along aerodynamic optimization trajectories. Convergence analysis and controlled experiments examine how initialization quality affects subsequent numerical convergence and computational cost. We further demonstrate the practical value of the framework in aerodynamic shape optimization and extend the same coupling approach to three-dimensional flows over realistic wing configurations on multi-million-cell meshes.

## Results

### Method overview

Surrogate–Newton coupling uses learned full-field predictions to initialize a residual-controlled numerical solution. Figure 1a illustrates how a prediction can retain useful global flow structures under geometric distribution shift, even when its standalone accuracy is inadequate. Using this predicted field as the initial state can reduce the numerical work required to achieve a prescribed residual ratio compared with initialization from a uniform flow field. Figure 1b summarizes the resulting workflow, which comprises prediction, state transfer, and iterative correction.

For a given geometry and flow conditions, denoted by $\boldsymbol{\theta} = (\mathcal{G}, \mathbf{c})$, where $\mathbf{c}$ comprises the Mach number Ma, angle of attack $\alpha$, and Reynolds number Re, the steady state $\mathbf{U}^{\star}$ satisfies

$$\mathbf{R}(\mathbf{U}^{\star}; \boldsymbol{\theta}) = \mathbf{0}, \tag{1}$$

where $\mathbf{R}$ includes the governing-equation discretization and boundary treatment. Here, the target system comprises the coupled Reynolds-averaged Navier–Stokes (RANS) and Spalart–Allmaras (SA) equations, as implemented in ADflow.[33]

The surrogate supplies the complete flow and turbulence fields,

$$\widehat{\mathbf{U}}_{\mathrm{s}} = \mathcal{G}_{\mathrm{s}}(\boldsymbol{\theta}; \phi), \qquad \mathbf{U}^{(0)} = \mathcal{I}(\widehat{\mathbf{U}}_{\mathrm{s}}), \tag{2}$$

where $\phi$ denotes the learned parameters, and $\mathcal{I}$ maps the prediction to the solver representation and grid ordering. We use a flow-field Schrödinger bridge (FSB) with deterministic multistep inference and refer to the coupled implementation as FSB–NK. The same interface accommodates single-pass predictors.

During the NK phase, coupled state corrections are computed using a local linearization of the target residual. The inexact update takes the form

$$\mathbf{J}_k \mathbf{s}_k = -\mathbf{R}_k + \boldsymbol{\ell}_k, \qquad \mathbf{U}^{(k+1)} = \mathbf{U}^{(k)} + \lambda_k \mathbf{s}_k, \tag{3}$$

where $\mathbf{R}_k = \mathbf{R}(\mathbf{U}^{(k)}; \boldsymbol{\theta})$, $\mathbf{J}_k$ is the physical-state residual Jacobian, $\mathbf{s}_k$ is the computed update, $0 < \lambda_k \leq 1$ is the step length, and $\boldsymbol{\ell}_k = \mathbf{J}_k \mathbf{s}_k + \mathbf{R}_k$ is the effective defect relative to the target Newton equation.[34] The linearized system is solved approximately using ADflow's preconditioned Jacobian-free NK procedure.[24] Within each solver call, the numerical solver controls step acceptance and re-evaluates the residual after each accepted update.

Numerical convergence of the returned state, denoted by $\mathbf{U}_{\mathrm{NK}}$, is recorded when its residual ratio satisfies

$$r(\mathbf{U}_{\mathrm{NK}}; \boldsymbol{\theta}) = \frac{\|\mathbf{R}(\mathbf{U}_{\mathrm{NK}}; \boldsymbol{\theta})\|_2}{\|\mathbf{R}(\mathbf{U}_{\mathrm{uniform}}; \boldsymbol{\theta})\|_2} \leq \epsilon, \tag{4}$$

where $\mathbf{U}_{\mathrm{uniform}}$ denotes the case-specific uniform-flow initialization, and $\epsilon$ is the prescribed residual-ratio tolerance. We write $r_k = r(\mathbf{U}^{(k)}; \boldsymbol{\theta})$ throughout the iterations. Runs terminated by the time limit may return a state that does not satisfy this convergence criterion.

Production evaluations begin with approximate Newton–Krylov (ANK) iterations and switch to NK when the residual ratio $r$ falls below a prescribed threshold. If startup stalls or the residual rebounds, the current numerical field can be passed back to the FSB model for one re-prediction, followed by a restarted solve within the remaining time limit. Details of the predictor construction, convergence stabilization, and evaluation protocols are provided in the Methods section.

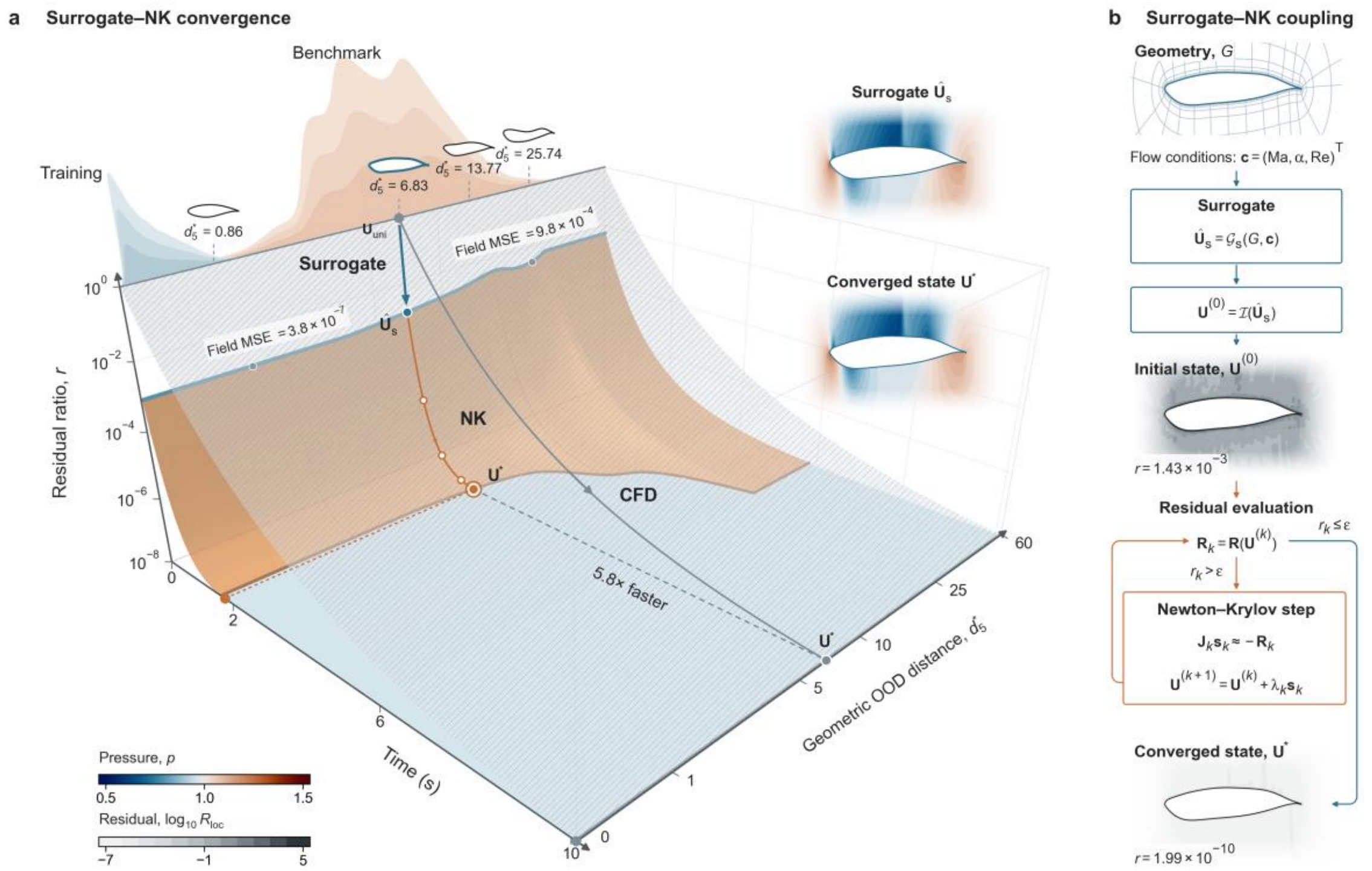

**Figure 1 | Surrogate–Newton coupling for steady CFD under geometric distribution shift. a,** Data-informed schematic of surrogate–Newton convergence. Surface interiors and Newton–Krylov (NK) node spacing are schematic. The residual ratio $r$ (equation (4)), elapsed time, and geometric out-of-distribution (OOD) distance share a common convergence frame. The geometric OOD coordinate $d_5^*$ is the mean distance to the five nearest training geometries, normalized by the training-set leave-one-out 99th percentile (Methods). The geometry distributions cover 27,117 supercritical training records and 3,405 benchmark geometries. Field mean-squared error (MSE) labels refer to the two gray points on the surrogate surface and represent the average five-channel error across the six available flow conditions per geometry. The CFD timing reference uses the pooled median residual-ratio history of 200 paired queries, whereas surrogate–NK endpoints use distance-bin median histories. The pale blue region shows the calibrated timing differences, and the dashed guide marks the 5.8-fold speedup at $r = 10^{-8}$. Pressure contours compare a surrogate prediction with the converged CFD reference for the same geometry and flow conditions. Blue denotes surrogate initialization, copper denotes NK correction, and gray denotes uniform-initialized CFD. **b,** Geometry and flow conditions are provided to the surrogate, and its prediction is transferred to the target solver. Residual evaluation, inexact Krylov solution, and damped updates constitute the NK loop (equations (2)–(4)). Initial and corrected fields show the local residual magnitude $R_{\mathrm{loc}}$ on a common logarithmic color scale. The loop terminates at $r_k \le \epsilon$, although budget-limited runs may return earlier.

## Convergence analysis

Classical Newton estimates bound the error by $E_{k+1} \le C E_k^2$ using worst-case constants across all perturbation directions.[35,36] For the stiff RANS–SA systems considered here, strong flow–turbulence coupling across disparate spatial scales makes such isotropic bounds overly conservative for assessing convergence from predicted initial states. Because the inverse-Jacobian-mapped nonlinearity varies strongly with direction, neural prediction errors need not excite the most amplified modes. We therefore develop a directional local-contraction framework based on the actual prediction error and its evolution under Newton correction.

For idealized full-step Newton corrections with exact linear solves, let $E_k$ denote the state-error norm after $k$ corrections in a fixed nondimensionalized norm $\|\cdot\|_X$. At the initial state $\mathbf{U}^{(0)}$, define $\mathbf{e}_0 = \mathbf{U}^\star - \mathbf{U}^{(0)}$ and $E_0 = \| \mathbf{e}_0 \|_X$. Mapping the nonlinear defect of the residual linearization through the inverse Jacobian determines the error remaining after the first correction and the associated directional-correctability margin:

$$\mathbf{g}_0 = \mathbf{J}_0^{-1}(\mathbf{R}_0 + \mathbf{J}_0\mathbf{e}_0), \qquad \chi_0 = 1 - \frac{\|\mathbf{g}_0\|_X}{E_0}. \tag{5}$$

Thus, $E_1 = (1 - \chi_0)E_0$, so $\chi_0 > 0$ indicates that the local linearization reduces the initial error. First-step error reduction alone does not guarantee convergence, however, because subsequent iterations encounter updated states and error directions.

To bound the nonlinear response along subsequent steps, we construct an initial directional scale $S_0$ from finite probes along $\mathbf{e}_0$ and $\mathbf{g}_0$ at $\mathbf{U}^{(0)}$, taking the larger response normalized by squared probe amplitude (Methods). A fixed multiplier $\beta \geq 1$ accounts for growth of this directional response as the state and error direction evolve.

**Assumption 1. Bounded directional response.** Along the generated Newton sequence, as long as the errors through the current iterate do not exceed $E_0$, the normalized nonlinear response remains bounded by $\beta S_0$, giving

$$E_{k+1} \leq \beta S_0 E_k^2, \qquad k \geq 1.$$

Whether this directional bound yields contraction depends on the error remaining after the first correction, quantified by the conditional convergence margin

$$\kappa_0 = 1 - \beta S_0 \| \mathbf{g}_0 \|_X. \tag{6}$$

A positive $\kappa_0$ places the post-first-step error $\| \mathbf{g}_0 \|_X$ strictly below the local contraction scale $1/(\beta S_0)$.

**Proposition 1. Conditional convergence and error bound.** Under Assumption 1 and the state-admissibility and Jacobian-invertibility conditions detailed in Methods, $\kappa_0 > 0$ guarantees that the idealized sequence remains within the initial error range and converges to $\mathbf{U}^\star$, with

$$E_K \leq E_0(1 - \chi_0)(1 - \kappa_0)^{K-1}, \qquad K \geq 1. \tag{7}$$

Equation (7) separates the initial error magnitude $E_0$, the first-step error reduction $(1 - \chi_0)$, and the subsequent contraction factor $(1 - \kappa_0)$, linking the utility of a predicted state to both its accuracy and its directional compatibility with Newton correction. For diagnostic evaluation across initializations, we calibrated $\beta \simeq 1.553$ on independent NK trajectories for 100 geometries and held it fixed without recalibration (Methods). Because practical workflows also involve damping, inexact Krylov solves, and ANK startup, $\chi_0$ and $\kappa_0$ serve as testable initial-state diagnostics rather than unconditional a priori guarantees for the production solver. Below, we first establish the optimization-derived OOD benchmark and macroscopic recovery performance (Figs. 2 and 3), and then directly evaluate how these directional margins explain observed NK convergence in controlled experiments (Fig. 4). Detailed definitions, regularity conditions, proofs, and the inexact-update error identity are provided in Methods.

## Out-of-distribution evaluation on an optimization-derived transonic aerodynamic benchmark

### *A geometrically shifted benchmark derived from aerodynamic optimization*

The benchmark includes 16,399 transonic flow cases across 3,405 airfoil geometries drawn from aerodynamic optimization trajectories (Fig. 2a,c). All cases have converged RANS reference solutions.

The benchmark exhibits a systematic geometric shift relative to the supercritical training population used as the reference in Fig. 2. In terms of the normalized mean distance to the five nearest training neighbors, $d_5^*$, 99.6% of benchmark geometries exceed the training 99th-percentile (P99) threshold. The population median increases from 0.24 in the training set to 6.90 in the benchmark (Fig. 2b). The first three principal components from principal component analysis (PCA), fitted to the supercritical training geometries, capture 92.7% of the training variance but only 27.2% of the benchmark variance. In the representative RAE2822 optimization history, elite designs exceed the supercritical training P99 distance threshold, whereas some intermediate candidates lie substantially farther from the training population than the retained elite designs (Fig. 2c). These findings highlight the necessity for reliable flow-field and aerodynamic evaluation across the broader geometric range explored by intermediate candidates.

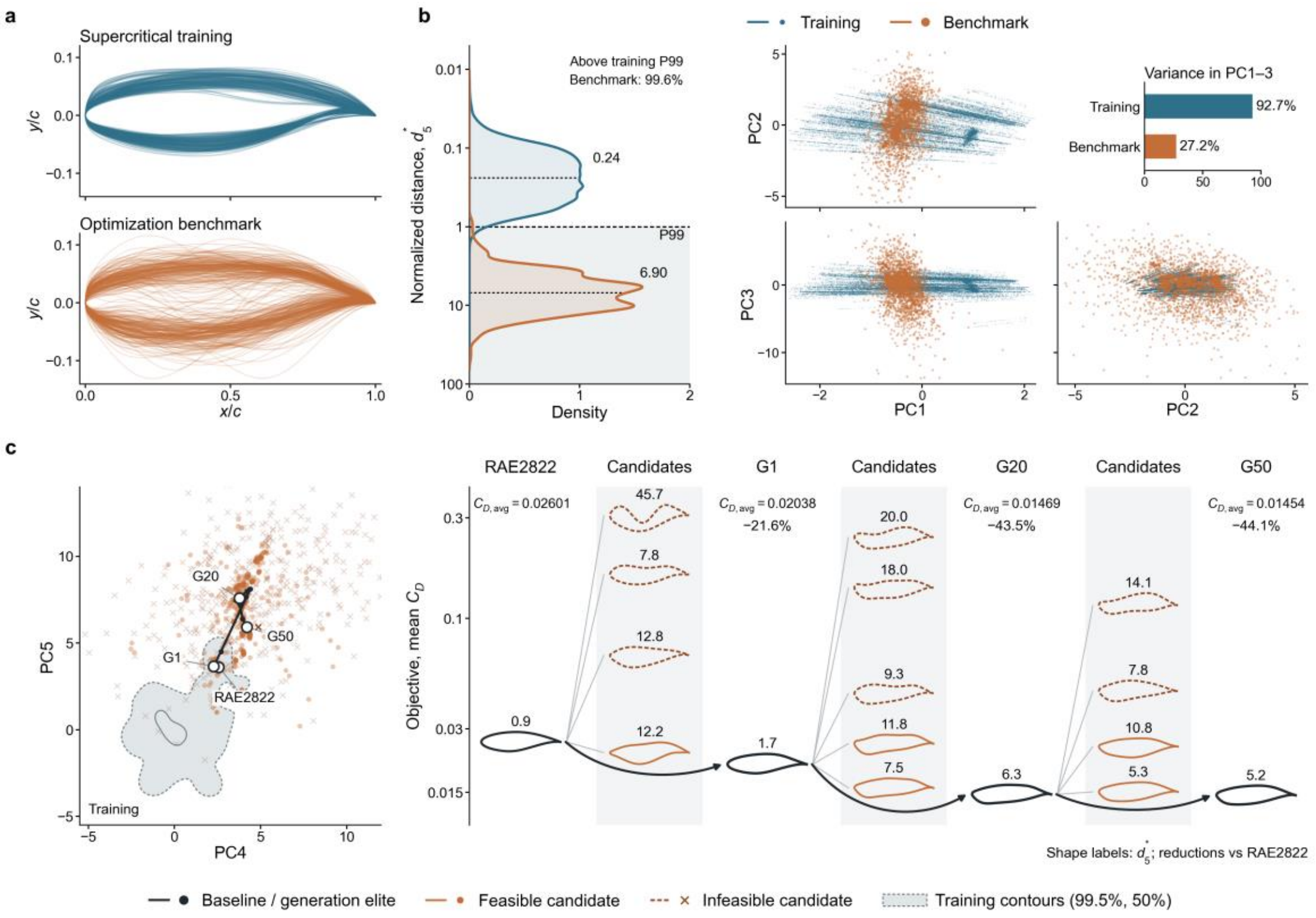

**Figure 2 | Optimization-derived airfoils define a geometrically shifted benchmark for transonic aerodynamic design. a,** Representative supercritical training and optimization-benchmark airfoils, shown with chord-normalized coordinates and twofold vertical scaling. **b,** Geometry distributions relative to the 27,117 supercritical training geometry records. The distance $d_5^* = d_5/P_{99,\mathrm{train}}$ is defined as the mean root-mean-square (RMS) wall-coordinate distance to the five nearest training neighbors, normalized by the training-set leave-one-out P99. The left panel shows the density per unit $\log_{10}$ distance against a logarithmic vertical distance axis. Blue and copper denote the training and benchmark populations, respectively. The horizontal dashed line and gray shading mark $d_5^* \geq 1$; black guides indicate the medians, and the annotation reports the percentage of the benchmark population above P99. Pairwise PC1–PC3 projections use the principal component analysis (PCA) basis fitted to the training geometries. The bars show the fractions of the total geometric variance in each population captured by these three principal components. **c,** CFD-driven RAE2822 optimization history. Left, PC4–PC5 projections of evaluated candidates highlight geometric variations not represented by PC1–PC3. Circles and crosses denote feasible and infeasible candidates, respectively. The black line traces the available elites, with the baseline and generations G1, G20, and G50 labeled. Gray contours enclose approximately 50% and 99.5% of the training projections. PCA scores are expressed in training-standard-deviation units. Right, selected candidate and elite geometries. Solid copper and dashed dark-copper outlines denote feasible and infeasible candidates, respectively; black outlines denote the baseline and elites. Vertical positions represent the mean drag objective $C_{D,\mathrm{avg}}$ on a logarithmic axis, while horizontal positions indicate generation order. Headings report elite objective values and their reductions relative to RAE2822, and outline labels give $d_5^*$. Thin links illustrate exploration, whereas thick arrows connect selected elites; neither implies parent–child lineage.

## *High-accuracy flow evaluation under geometric distribution shift*

Across 16,399 flow cases from the geometrically shifted benchmark, FSB–NK achieves orders-of-magnitude improvements in flow-field and aerodynamic accuracy over standalone FSB predictions, as measured against CFD reference solutions (Fig. 3a; Methods). The median field mean-squared error (MSE) decreases from $2.5 \times 10^{-5}$ to $9.8 \times 10^{-12}$, corresponding to a reduction of more than six orders of magnitude. Improvements also extend to the high-error tail of aerodynamic predictions: the 95th-percentile absolute lift-coefficient error decreases from $5.6 \times 10^{-2}$ to $3.6 \times 10^{-5}$, while the corresponding drag-coefficient error decreases from $4.7 \times 10^{-3}$ to $2.4 \times 10^{-6}$, with both reductions exceeding three orders of magnitude. With the solver-time budget set to 12 s per case, FSB–NK returned solutions meeting the residual target of $10^{-8}$ in 15,866 cases (96.75%).

Among the 812 cases that triggered re-prediction, the convergence rate rose from 70.9% with re-prediction disabled to 84.6% with it enabled. The 90th percentile of field MSE fell from $2.12 \times 10^{-3}$ to $6.18 \times 10^{-4}$.

Representative benchmark geometries and an interpolation path further examine how these accuracy gains extend with increasing geometric shift. Across six benchmark geometries ordered by increasing normalized geometric distance, beginning with RAE2822, FSB captures the broad pressure trends, whereas FSB–NK closely reproduces the CFD pressure and skin-friction profiles, with particularly pronounced improvements in $C_f$ accuracy (Fig. 3b). An interpolation path from the same starting geometry to a thick ellipse tests stronger geometric extrapolation (Fig. 3c). Although standalone FSB predictions become progressively less accurate toward the ellipse, they remain effective initial states for numerical correction along the sampled path.

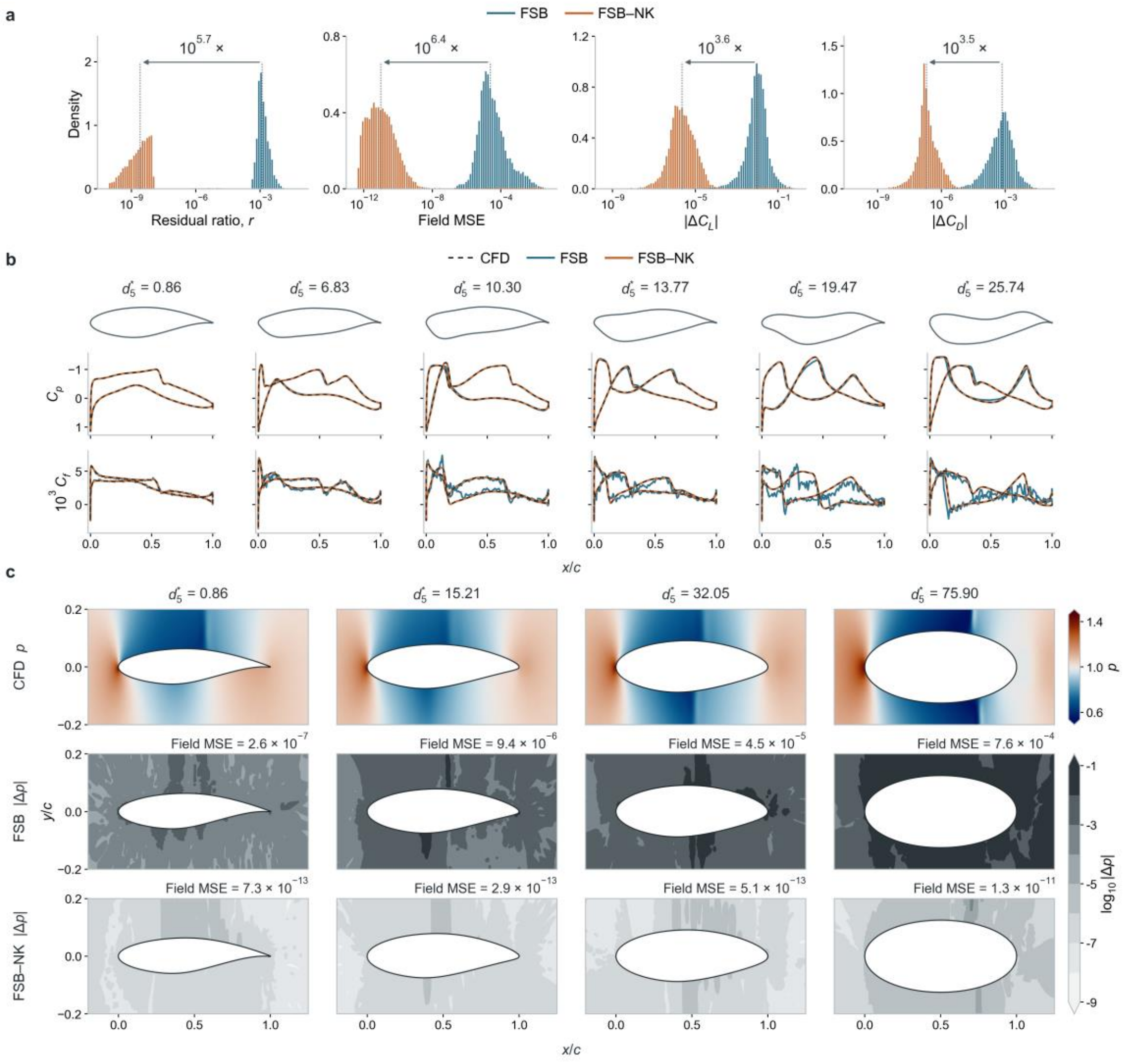

**Figure 3 | Surrogate–Newton coupling delivers high-accuracy flow evaluation under geometric distribution shift. a,** Probability densities of the residual ratio $r$, field mean-squared error (MSE), and absolute lift- and drag-coefficient errors for standalone flow-field Schrödinger bridge (FSB) predictions and returned FSB–NK fields across the 16,399-case benchmark. The residual distribution uses cases with native ratios available for both methods. Each metric uses equal-width bins in $\log_{10}(x)$, shared by both methods; density represents the fraction of cases per unit $\log_{10}(x)$. Arrows connect the method medians and indicate their FSB-to-FSB–NK ratios. Each metric has an independent vertical range. Field MSE is calculated by averaging squared errors across the five physical channels; field and aerodynamic errors are evaluated against CFD references. **b,** Six benchmark geometries ordered by increasing $d_5^*$, starting with RAE2822. Rows show the airfoil geometries, surface-pressure coefficient $C_p$, and skin-friction coefficient $C_f$ for CFD, FSB, and FSB–NK results. Geometries are shown with twofold vertical scaling. **c,** Four geometries interpolating from the same RAE2822 case to an ellipse of thickness-to-chord ratio 0.25. Rows show the CFD pressure $p$ and the absolute pressure errors for FSB and FSB–NK, displayed on a common $\log_{10}|\Delta p|$ scale. At the ellipse endpoint, the Field MSE for pressure decreases from $7.6 \times 10^{-4}$ to $1.3 \times 10^{-11}$ after correction. Cases in b and c use $\mathrm{Ma} = 0.74$, $\alpha = 1^\circ$, and $\mathrm{Re} = 1.6378 \times 10^7$.

## Mechanistic analysis of surrogate–Newton coupling

Comparable initial residuals do not necessarily imply equally effective initializations for NK convergence. At matched residual levels, FSB initializations have a median scaled state MSE ($E_0^2$) of $1.80 \times 10^{-4}$, compared with $7.65 \times 10^{-3}$ for fields obtained through pseudo-time marching. Thus, the learned predictions are closer to the target state even before residual convergence is achieved. Controlled comparisons in a representative case distinguish the contributions of initialization quality and solver choice. With the NK settings fixed, the FSB initialization reaches the prescribed tolerance much sooner than the residual-matched pseudo-time field. Conversely, with the FSB initial field held fixed, NK reaches the same tolerance much faster than pseudo-time marching (Fig. 4a). These controls demonstrate the complementary contributions of the two components: the surrogate provides an informative full-field initialization, while NK efficiently converts this initial state into a converged solution.

Convergence analysis connects these initialization differences to state-error magnitude and the directional response of the residual linearization. The margins $\chi_0$ and $\kappa_0$ distinguish initial directional correctability from the stronger requirement for convergence under Assumption 1. Uniform initializations illustrate this distinction: 98% have $\chi_0 > 0$, yet none have $\kappa_0 > 0$ (Fig. 4b). Thus, an initial linearization that reduces the error in an idealized first correction does not necessarily place the resulting state within the error scale required by the conditional convergence criterion. Positive $\kappa_0$ occurs for 62% of FSB initializations, compared with 9% of residual-matched pseudo-time states and 0% of uniform states. The lower state errors of learned initializations are therefore accompanied by a higher prevalence of directional responses satisfying the conditional convergence criterion, helping explain their advantage in the controlled solver comparisons.

Learning quality continues to influence the cost of convergence and the variability of outcomes within the coupled workflow. Across geometric-distance bins, the median $\chi_0$ remains positive, whereas the median $\kappa_0$ falls below zero in the more distant bins (Fig. 4c). A non-positive $\kappa_0$ indicates that the sufficient convergence condition is not satisfied; however, it does not necessarily imply failure of numerical correction. Some difficult queries retain larger residuals within the same solver-time budget, with the mean FSB–NK residual ratio increasing from $2.1 \times 10^{-8}$ to $2.4 \times 10^{-5}$. Increasing the training set size reduces the solver work and the time required to reach the same convergence threshold (Fig. 4d). With approximately 1,000 training samples (1% of the full dataset), the population median residual ratio reaches $10^{-8}$ after 2.90 s of solver time, compared with 10.10 s for uniform-initialized CFD. The full-data model further reduces this time to 1.76 s. With nearest-neighbor (1-NN) initialization from a training CFD field, the median residual ratio reached the same target in 9.22 s (Fig. 4d).

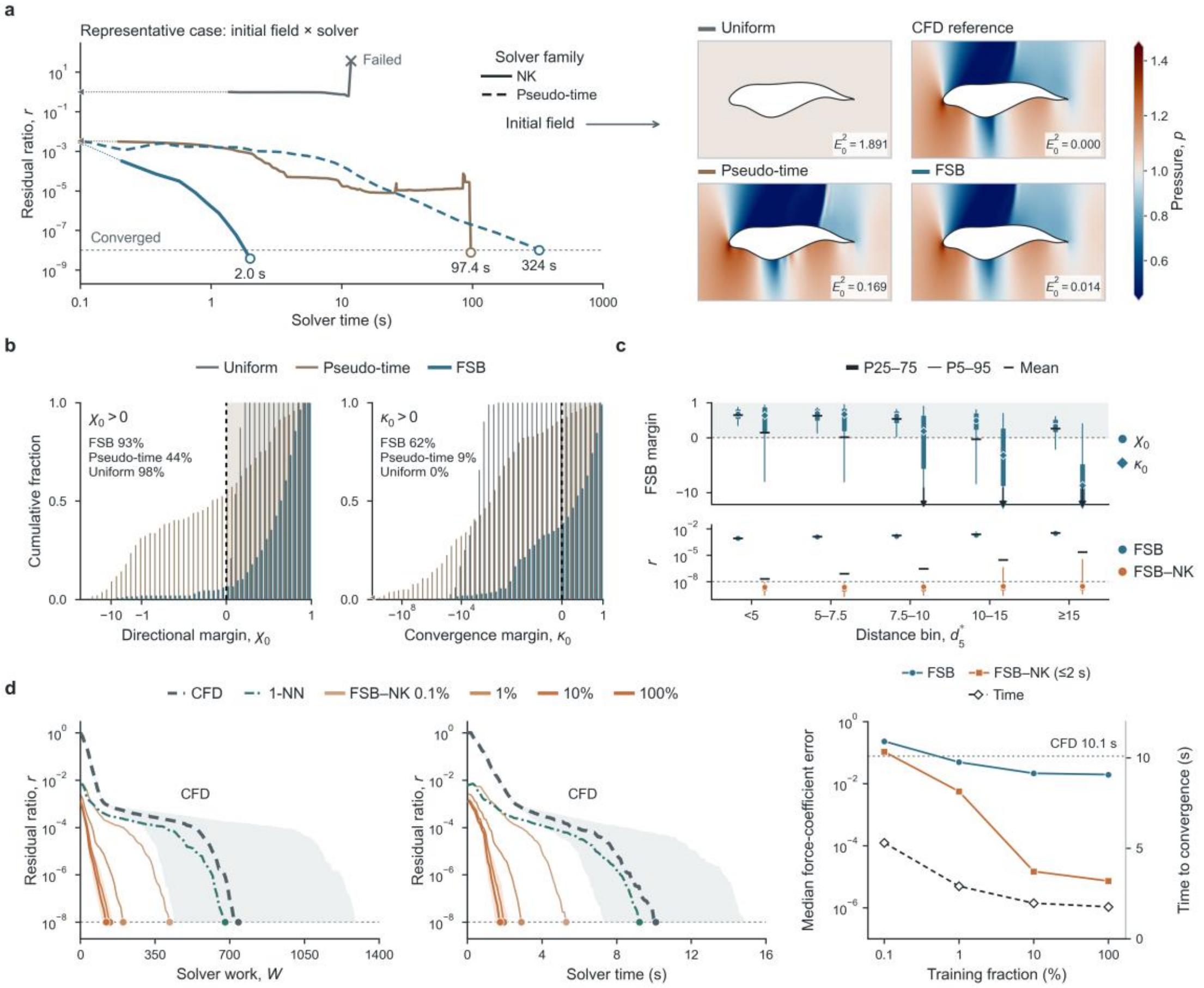

**Figure 4 | Learned initialization and Newton–Krylov correction provide complementary convergence benefits. a,** Controlled comparisons at $\mathrm{Ma} = 0.76$, $\alpha = 5.496^\circ$, and $\mathrm{Re} = 1.6821 \times 10^7$. Left, residual-ratio trajectories, with color denoting the initialization method and line style distinguishing NK correction from pseudo-time marching. Right, initial pressure fields and the converged CFD reference; $E_0^2$ denotes the scaled full-state mean-squared error (MSE). Flow-field Schrödinger bridge (FSB) and pseudo-time initializations have comparable residual ratios near $3 \times 10^{-3}$ and use the same Newton–Krylov (NK) settings. **b,** Empirical cumulative distributions of the initial directional-correctability margin $\chi_0$ and conditional convergence margin $\kappa_0$ for FSB, uniform, and residual-matched pseudo-time initializations from a study that sampled 200 queries. Thin bars sample the distributions on symmetric-logarithmic axes. Positive $\chi_0$ indicates error reduction in the first idealized Newton correction, whereas positive $\kappa_0$ is the sufficient convergence criterion under Assumption 1. Shading indicates positive margins, and annotations report the corresponding population fractions. **c,** Dependence on $d_5^*$. Top, initial FSB margins from the diagnostic study, plotted on symmetric-logarithmic axes. Bottom, residual ratios before and after NK correction for paired cases. Symbols indicate medians; thick and thin intervals span the 25th–75th (P25–75) and 5th–95th (P5–95) percentiles, respectively; black ticks indicate arithmetic means. Downward arrows identify means outside the plotted range, and triangles indicate truncated intervals. **d,** Effect of training-set fraction on 200 matched benchmark cases. Left and middle, median residual ratios plotted against solver work $W$ (minor iterations) and time, respectively, for the complete approximate Newton–Krylov (ANK)-to-NK solve. Copper curves show FSB–NK at the indicated training fractions, gray curves show uniform-initialized CFD, and teal dash-dotted curves show the 1-NN baseline. Bands span the P25–75 intervals for CFD and full-data FSB–NK. Right, median force-coefficient error $E_F = |\Delta C_L| + 10|\Delta C_D|$ for initial FSB and FSB–NK within a 2 s solver budget (left axis), and time to convergence at $r = 10^{-8}$ (right axis).

## Reliable and efficient aerodynamic design optimization

We next evaluated surrogate–Newton coupling in aerodynamic design optimization, extending the assessment beyond offline flow evaluation to applications closer to engineering practice. We considered two design tasks: a classical multipoint transonic drag-minimization problem and a lift-to-drag ratio maximization problem for a Mars unmanned aerial vehicle (UAV). Optimization driven by standalone FSB, FSB–NK, and CFD was compared across both tasks.

In the transonic task, standalone FSB guided the search toward plausible supercritical profiles within the design regime represented by its training data. Coupling FSB with NK brought the aerodynamic estimates into closer agreement with independent CFD evaluations: across the reviewed candidates, the mean absolute drag discrepancy decreased from 2.92 to 0.24 counts (Fig. 5a). This improved agreement was accompanied by better verified designs, with a mean drag 14.0 counts lower than that obtained through standalone FSB optimization and 0.3 counts lower than that obtained through CFD-driven optimization. Accurate evaluation was also important when selecting candidates near constraint boundaries. For one candidate, the FSB prediction of $C_m = -0.09181$ satisfied the lower bound of $-0.092$ for the pitching moment, whereas independent CFD re-evaluation yielded $C_m = -0.09226$, reversing the feasibility assessment. FSB–NK achieved these improvements in evaluation agreement and verified design performance while delivering a 6.7-fold speedup over CFD in optimization time (Fig. 5a,b).

The Mars-UAV task[37] provided an extreme test of generalization, involving thin, highly cambered airfoils and low-Reynolds-number operating conditions representative of Martian flight, well beyond the FSB training distribution. Standalone FSB produced unphysical force estimates, including a negative drag coefficient of $C_D = -0.156$ for one candidate, and could no longer provide reliable aerodynamic evaluations. Post hoc CFD re-evaluation recovered acceptable designs from four of the six standalone FSB runs. In contrast, FSB–NK produced designs that passed independent CFD verification in all six runs, with a mean peak lift-to-drag ratio of 27.88, close to the value of 28.09 obtained through CFD-driven optimization (Fig. 5c,d).

We further developed a more efficient optimization strategy based on Top-$K$ screening, in which surrogate predictions are used to rank candidates and NK correction is restricted to the top $K$ candidates in each generation. In the transonic task, Top-4 delivered a 17.3-fold speedup over CFD in optimization time (Fig. 5a). The mean optimization time decreased from 14.12 to 0.82 h, whereas the mean verified drag was 2.7 counts higher than that obtained with full correction. In the Mars-UAV task, Top-4 achieved a mean verified peak lift-to-drag ratio of 26.80, compared with 27.88 for full correction, while providing an estimated 5.5-fold speedup over CFD-driven optimization (Fig. 5c).

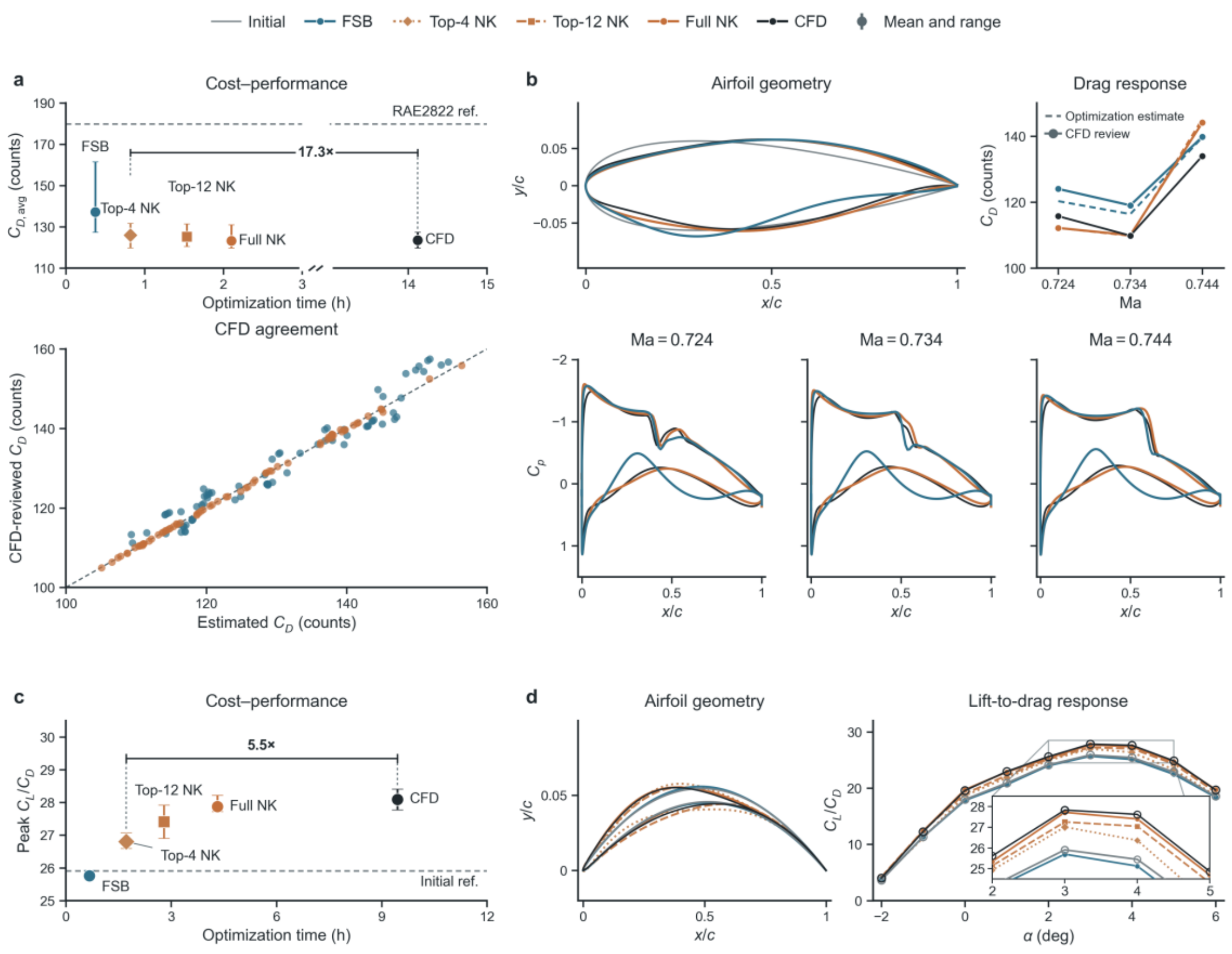

**Figure 5 | Aerodynamic design optimization with surrogate–Newton coupling. a,** Multipoint transonic drag minimization starting from NACA 0012 at Ma $= 0.724, 0.734, 0.744$ and targeting $C_L = 0.824$, across seeds 0–3. Top, three-condition mean drag, independently verified by CFD, plotted against optimization time. The time axis is broken between 3 and 13 h; the bracket indicates the 17.3-fold ratio of mean CFD to Top-4 optimization time. Bottom, condition-wise comparison of optimization estimates with independent CFD re-evaluations across reviewed candidates. One drag count corresponds to $10^{-4}$ in $C_D$. **b,** Initial and selected geometries, drag responses, and surface-pressure profiles for transonic optimization seed 0. All $C_p$ profiles are independently evaluated by CFD; dashed and solid drag curves represent optimization estimates and CFD re-evaluations, respectively. **c,** Lift-to-drag maximization for a Mars unmanned aerial vehicle (UAV) at Ma $= 0.6$ and Re $= 20{,}000$, across six seeds. Performance is defined as the maximum independently CFD-re-evaluated $C_L/C_D$ across nine angles of attack, ranging from $-2^{\circ}$ to $6^{\circ}$. The horizontal axis represents estimated optimization time, excluding independent verification. FSB performance is based on four accepted designs, whereas the reported time includes all six runs. The bracket indicates the 5.5-fold ratio of mean estimated optimization times for CFD and Top-4. **d,** Initial and selected geometries for Mars-UAV optimization seed 2, together with independently CFD-re-evaluated lift-to-drag responses; inset, enlarged view of the peak region. Geometries are shown with approximately 6.2-fold vertical scaling. In b and d, colors identify the optimization method used to select each design. Full NK applies FSB–NK to every candidate, whereas Top-4 and Top-12 apply NK correction to the top four and twelve candidates, respectively, as ranked by the surrogate in each generation. Cost–performance markers and error bars in a and c indicate seed means and minimum–maximum ranges, respectively. Gray dashed reference lines indicate RAE2822 in a and the initial section in c.

## Extending surrogate–Newton coupling to three-dimensional wing flows

We extended surrogate–Newton coupling to three-dimensional wing flows on meshes containing approximately 3.5 million cells, using a three-dimensional FSB predictor to provide the complete flow and turbulence fields. The paired benchmark comprises 1,085 flow cases across 942 wing geometries derived from sectional-shape and planform optimization tasks. The median $d_5^*$ value is 1.57 for the section-optimization population and 3.82 for the planform-optimization population, with 99.8% and 89.0% of the respective geometries exceeding $d_5^* = 1$ (Fig. 6a). FSB–NK reached a residual ratio of $10^{-6}$ in 1,081 of the 1,085 paired wing flow cases (99.63%).

In pilot tests on three-dimensional wings, 1-NN initialization had a markedly lower convergence rate than uniform initialization. When it converged, 1-NN required more solver work and took substantially longer than uniform-initialized CFD. This may reflect sparser sampling of wing geometries, which makes it harder to match both geometry and operating conditions. We therefore use uniform-initialized CFD as the primary baseline.

Full-field initialization substantially reduces the computational cost of obtaining a converged three-dimensional solution. Across 60 matched cases in which both methods converged, the median elapsed times are 472 s for uniform-initialized multigrid CFD and 89 s for FSB–NK. The median paired speedup is 5.2-fold (Fig. 6b). The convergence histories indicate that a substantial portion of the baseline solver work is spent developing the flow on coarser grids, whereas FSB–NK begins directly on the finest grid using the predicted volume state.

Across all 1,085 wing flow cases, FSB–NK with fallback to the original FSB prediction reduced the median Field MSE by a factor of 349 and the median weighted lift-and-drag error by a factor of 35 relative to standalone FSB, using the original archived CFD solutions as references (Fig. 6c). In two separate examples from section and planform optimization, FSB captures the broad pressure distributions, whereas FSB–NK reduces surface-pressure errors and more closely reproduces the CFD reference profiles at the sampled spanwise stations (Fig. 6d).

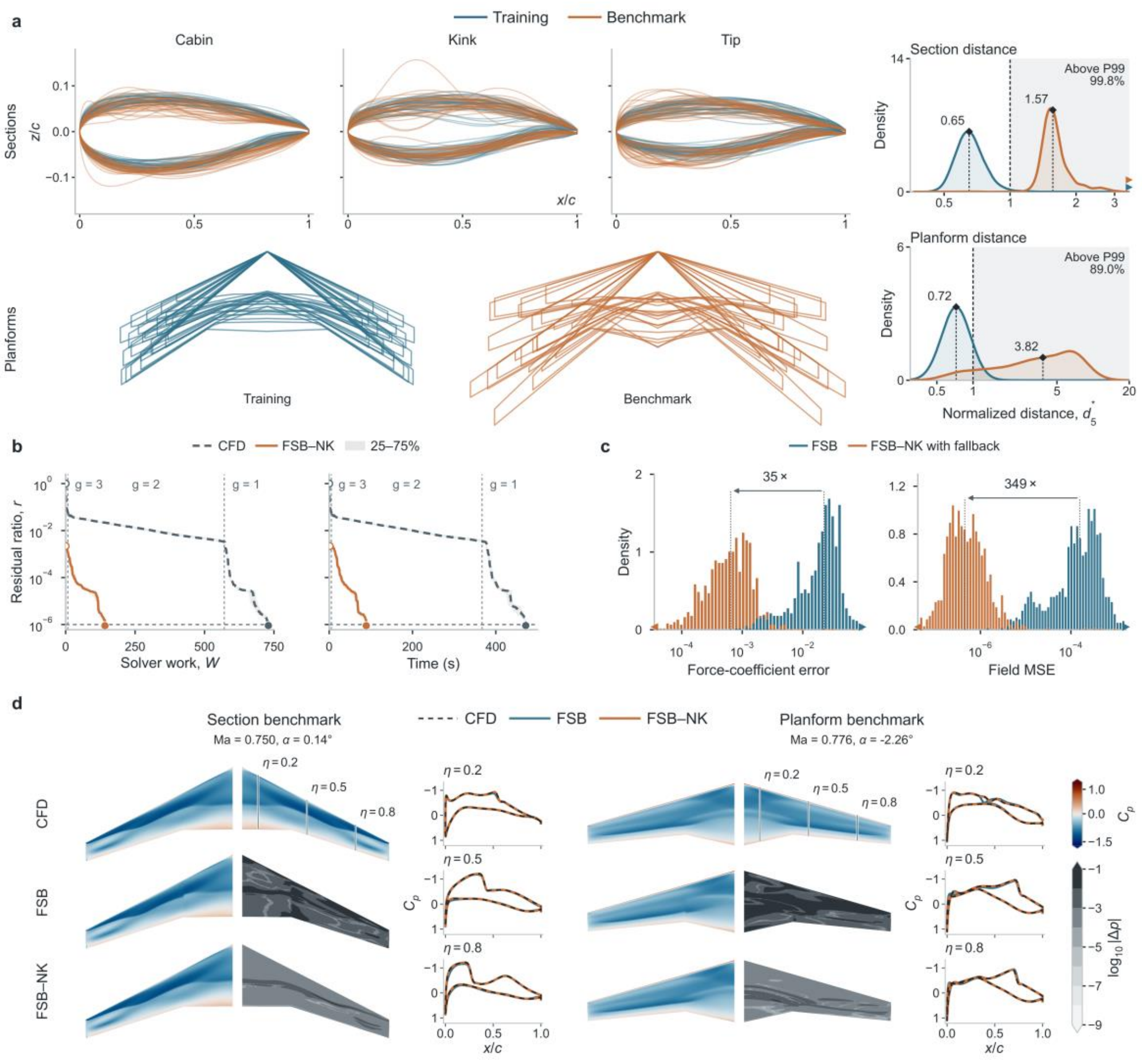

**Figure 6 | Surrogate–Newton coupling improves accuracy and reduces solution cost for three-dimensional wing flows. a,** Representative wing sections, planforms, and corresponding $d_5^*$ distributions for training geometries (blue) and optimization-benchmark geometries (copper). Section profiles are shown with twofold vertical scaling, whereas planforms share a common physical scale. Density is expressed per unit $\log_{10}$ distance. Dashed lines and shaded regions mark $d_5^* \geq 1$; dotted guides and diamonds indicate medians. Annotations report the fractions of benchmark geometries above P99. Boundary triangles indicate distribution tails extending beyond the plotted range. **b,** Residual-ratio histories plotted against solver work and elapsed time for 60 matched cases in which both methods reached the residual target. Dashed gray and solid copper lines represent the median histories for CFD and FSB–NK, respectively; shaded bands span the P25–75 intervals. Vertical guides indicate CFD grid transfers, whereas FSB–NK starts directly on the finest grid. Coarse-grid reductions are mapped continuously onto the finest-grid scale (Methods). Finest-grid residual ratios use a common uniform-state reference for each case and a $10^{-6}$ target. **c,** Force-coefficient error, $E_F = |\Delta C_L| + 10|\Delta C_D|$, and Field MSE for pressure across all 1,085 cases, evaluated against the corresponding original archived CFD solutions. For cases that did not meet the residual target, the FSB–NK workflow is scored using the original FSB prediction. Log-binned densities show the error distributions, with boundary triangles indicating tails extending beyond the plotted range. Arrows connect the method medians and indicate the corresponding FSB-to-FSB–NK ratios. **d,** Separate section- and planform-optimization examples. CFD results show the surface $C_p$ on both mirrored semispans, whereas FSB and FSB–NK results show $C_p$ on the left and $\log_{10}|\Delta p|$ relative to matched CFD results on the right. Curves represent distributions sampled at the indicated normalized semispan stations $\eta = 0.2, 0.5, 0.8$.

## Discussion

The computational value of a learned flow field extends beyond its accuracy as a standalone prediction. Under geometric distribution shift, a prediction may preserve useful global flow structure while lacking the precision required for aerodynamic assessment. This structure gives Newton–Krylov iterations an informed starting state, and residual-controlled solution supplies the numerical accuracy needed for evaluation. The residual-matched comparisons demonstrate why the initial residual alone is inadequate: similar residuals can conceal substantial differences in state errors and convergence behavior. The directional analysis further relates these differences to the magnitude of the state error and the local Newton response along the prediction-error direction.

Beyond offline benchmarks, the optimization experiments show that surrogate–Newton coupling supports aerodynamic design beyond the effective range of standalone predictions, with selected designs verified through independent CFD evaluations, while the three-dimensional wing results demonstrate scalability to multi-million-cell meshes. Learning remains central to this coupled computation.[38,39] In the two-dimensional study, increasing the training data reduces subsequent solver work, and even 1% of the training set (approximately 1,000 samples) lowers the median time to convergence from 10.10 s to 2.90 s. In many-query workflows such as aerodynamic optimization or parametric design sweeps, this data efficiency enables offline data-generation and training costs to be amortized across repeated high-fidelity evaluations. These findings suggest assessing learned flow models through the convergence cost and reliability they enable, alongside their standalone prediction accuracy.[40,41]

Numerically converged solutions still inherit the turbulence-modeling and discretization errors of the underlying RANS solver. In addition, both our two- and three-dimensional studies adopt fixed structured-mesh topologies to preserve computational efficiency and limit GPU memory footprint during full-field inference. Crucially, however, this choice is an implementation trade-off rather than a methodological restriction: because the surrogate interacts with the numerical solver solely through full-field state initialization, the coupling is modular and agnostic to both network architecture and mesh topology. Extending the framework to unstructured or adaptive mesh topologies, or to other stiff steady-state or implicit PDE solvers, would require a predictor that supplies an initial state acceptable to the target solver, along with compatible state transfer and effective preconditioning on the target mesh.[42,43]

# Methods

## Flow formulation

Reference flow fields and numerical corrections are obtained using ADflow to solve the coupled steady compressible Reynolds-averaged Navier–Stokes and Spalart–Allmaras equations.[33,44] The target residual in equation (1) includes the spatial discretization and boundary treatment. Body-fitted meshes are generated using pyHyp.[45] Two-dimensional airfoil states contain $(\rho, u, v, p, \tilde{\nu})$ and are represented on a common cell-centered O-grid comprising $84 \times 304$ control volumes. Three-dimensional wing states contain $(\rho, u, v, w, p, \tilde{\nu})$ and are represented on a fixed 13-block topology comprising approximately 3.5 million cells. The flow and turbulence fields follow ADflow's solver nondimensionalization; in particular, $p$ is defined as the ratio of pressure to freestream pressure.

## Surrogate prediction of flow fields

### *Surrogate formulation*

The surrogate $\mathcal{G}_{\mathrm{s}}(\boldsymbol{\theta}; \phi)$ predicts the complete flow and turbulence fields from the geometry $\mathcal{G}$ and operating conditions $\mathbf{c}$. The same prediction is used for standalone aerodynamic evaluation or as the initial state for NK. The inputs comprise geometry descriptors, Mach number, angle of attack, and $\log_{10}\mathrm{Re}$, together with physical cell-center coordinates and structured-grid indices, collectively denoted by $X$. We implement the surrogate using an FSB, with a diffusion-transformer (DiT) endpoint network for airfoils[46] and a hierarchical multiblock Fourier neural operator (HM-FNO) for wings.[3]

### *Schrödinger bridge for flow fields*

FSB learns the steady-state endpoint of a bridge between a converged state and its case-specific uniform initialization.[47] Denoting these endpoints by $\mathbf{x}_0$ and $\mathbf{x}_1$, respectively, a training bridge state is given by

$$\mathbf{x}_t = w_0(t)\mathbf{x}_0 + w_1(t)\mathbf{x}_1 + \eta_{\mathrm{br}}\boldsymbol{\Sigma}_t^{1/2}\boldsymbol{\epsilon}, \qquad \boldsymbol{\epsilon} \sim \mathcal{N}(\mathbf{0}, \mathbf{I}),$$

where $\eta_{\mathrm{br}}$ sets the bridge noise amplitude, and the Gaussian-bridge weights and covariance connect the steady-state endpoint at $t = 0$ to the uniform endpoint at $t = 1$, with $w_0 + w_1 = 1$. The endpoint network $F_{\phi}(\mathbf{x}_t, t; X, \mathcal{G}, \mathbf{c})$ predicts $\mathbf{x}_0$ from the bridge state. In two dimensions, the DiT processes the bridge state and spatial inputs, incorporating geometry, operating-condition, and bridge-coordinate conditioning, to predict all five steady-state field channels.

Inference is deterministic ($\eta_{\mathrm{br}} = 0$). Starting from $\mathbf{x}^{(0)} = \mathbf{x}_1$, the predictor proceeds along decreasing bridge coordinates, $1 = t_0 > t_1 > \cdots > t_H \geq 0$ . Each endpoint prediction provides the input for the next bridge step:

$$\hat{\mathbf{x}}_0^{(j)} = F_\phi(\mathbf{x}^{(j)}, t_j; X, \mathcal{G}, \mathbf{c}), \qquad j = 0, \dots, H,$$

$$\mathbf{x}^{(j+1)} = w_0(t_{j+1})\hat{\mathbf{x}}_0^{(j)} + w_1(t_{j+1})\mathbf{x}_1, \qquad j < H.$$

The original uniform endpoint $\mathbf{x}_1$ remains fixed throughout the sequence. The final endpoint prediction $\hat{\mathbf{x}}_0^{(H)}$ provides the standalone flow field or the numerical initialization used in equation (2).

***Multiblock prediction for wing flows***

The three-dimensional endpoint network uses HM-FNO to predict all six flow and turbulence channels across the 13 mesh blocks.[3] The common structured-mesh topology enables efficient blockwise tensor operations and compact patch representations, thereby reducing GPU memory requirements compared with processing the entire mesh as a point cloud. This approach supports efficient full-field prediction on wing meshes containing millions of cells. Each block receives its bridge field $\mathbf{x}_{t,b}$ and spatial inputs $X_b$. Block-specific embeddings lift non-overlapping patches to a common feature space, and local Fourier operator layers shared across blocks generate features $h_b$. Geometry, operating conditions, and the bridge coordinate form a common conditioning vector $\zeta$. The local and global operations can be expressed schematically as

$$h_b = \mathcal{L}_\phi(\mathcal{E}_{\phi,b}(\mathbf{x}_{t,b}, X_b); \zeta), \qquad T = \mathcal{A}_\phi(\mathrm{concat}_{b=1}^{13} \mathcal{P}_\phi(h_b; \zeta); \zeta),$$

$$\hat{\mathbf{x}}_{0,b} = \mathcal{D}_{\phi,b}(h_b + \mathcal{C}_\phi(h_b, T); \zeta).$$

Here, $\mathcal{P}_\phi$ uses attention pooling within local windows to form compact tokens, with block identities and positional information added before concatenation. The global attention operator $\mathcal{A}_\phi$ aggregates information across tokens from all blocks. The shared cross-attention readout $\mathcal{C}_\phi$ enables each local patch to retrieve information from this global representation. Block-specific Fourier refinement and decoding, represented by $\mathcal{D}_{\phi,b}$, reconstruct the cell-level endpoint fields. Patch embedding uses replicated edge padding, which is removed after decoding. The decoded fields are then transformed back into physical variables and assembled in the solver's block and cell order. Interblock coupling and physical boundary conditions are evaluated through the target residual during numerical correction. The model follows the same bridge construction and deterministic endpoint-prediction sequence as the airfoil model.

***Surrogate training***

Let $Y$ denote the solver-nondimensional physical channels, with $Q = 5$ for airfoils and $Q = 6$ for wings. Training uses the following representation:

$$z_{aj} = \frac{h_a(Y_{aj}) - \mu_a}{\sigma_a}, \qquad h_a(y) = \begin{cases} \log(1 + k_v y), & a \text{ is the Spalart–Allmaras channel}, \\ y, & \text{otherwise}, \end{cases}$$

where $k_v > 0$ is a fixed turbulence-variable scale, and $\mu_a, \sigma_a$ represent the fixed training-set statistics of the transformed channels. The bridge endpoints and intermediate states are expressed using this representation. Inverting the transformation yields the physical prediction $\hat{Y}$, which is then converted into the solver state $\hat{\mathbf{U}}$ for residual evaluation and numerical initialization.

For both airfoils and wings, the field-reconstruction term compares the predicted steady-state endpoint with the reference field in standardized coordinates:

$$\mathcal{L}_{\text{field}} = \mathbb{E}[\frac{1}{QN_c} \sum_{j=1}^{N_c} \sum_{a=1}^{Q} w_j \, (\hat{z}_{0,aj} - z_{0,aj})^2].$$

Here, $N_c$ denotes the total number of cells, and the expectation is taken over training queries, bridge coordinates, and bridge noise. Airfoil reconstruction uses capped inverse-square-root cell-volume weights $w_j$, normalized to a unit mean to emphasize the smaller cells near the wall. Wing reconstruction uses $w_j = 1$, averaging over all six channels and all cells across the 13 blocks. Models are selected based on validation data and remain fixed throughout subsequent evaluation and optimization.

For airfoil accuracy assessment, the field MSE is calculated by averaging the squared errors equally across the five physical, solver-nondimensional channels and all cells, without applying training-statistic or cell-volume weighting. The MSE is evaluated after inverting the training transformation.

## Newton–Krylov method

### *Newton–Krylov iterations*

The predicted flow and turbulence fields provide the initial nonlinear iterate in equation (2). During the NK phase, ADflow applies the inexact updates defined in equation (3) to the coupled residual equations using the preconditioned generalized minimal residual (GMRES) method,[48] with Jacobian–vector products evaluated using directional residual differences and native nonlinear step control.[24,30,32] The residual is re-evaluated after each accepted update. Convergence is determined according to equation (4), with target values of $10^{-8}$ for the two-dimensional benchmark and transonic task, $10^{-10}$ for the Mars-UAV task, and $10^{-6}$ for wings.

### *Approximate Newton–Krylov startup*

Production solves begin with approximate Newton–Krylov (ANK) startup, followed by NK iterations. In the two-dimensional workflow, ANK switches to NK when the residual ratio relative to the case-specific uniform initial state falls below $10^{-4}$. The initial solve and any restarted solve share the prescribed solver-time budget; the two-dimensional benchmark sets this budget to 12 s per query. The solver checks the limit during iteration, so the actual return can take longer. In three dimensions, uniform-initialized CFD progresses through three grid levels, whereas FSB–NK starts directly on the finest grid. Both methods use the same finest-grid residual target, and the returned wing solutions undergo independent residual evaluation.

### *Re-prediction*

The two-dimensional calculations permit one FSB re-prediction when the ANK startup makes insufficient progress or the residual rebounds. Relative to the residual at the start of the solver call, re-prediction is triggered if the residual remains above 0.50 after five accepted ANK updates or above 0.30 after six accepted updates. A tenfold increase relative to the preceding or best recorded residual also triggers re-prediction, provided that the residual exceeds 0.01 relative to its value at the start of the solve.

At the trigger point, the current numerical field $\mathbf{x}_{\mathrm{num}}$ supplies an updated steady-endpoint estimate. This estimate is combined with the original uniform endpoint to construct a bridge input:

$$\mathbf{x}_{\mathrm{rp}} = w_0(t_{\mathrm{rp}})\mathbf{x}_{\mathrm{num}} + w_1(t_{\mathrm{rp}})\mathbf{x}_1, \qquad t_{\mathrm{rp}} = 20/999.$$

The endpoint network predicts a revised steady-state field from $\mathbf{x}_{\mathrm{rp}}$. This field initializes a new ANK-to-NK solve using the remaining solver-time budget, with further re-prediction disabled. The field obtained from the restarted solve is retained if it reduces the residual relative to that of the revised prediction and passes the physical-state and residual checks; otherwise, the revised prediction is returned. Numerical convergence of the returned state is assessed separately according to equation (4).

**Convergence analysis**

For a fixed query, let $\mathbf{U}^\star$ be the target root of $\mathbf{R}$ and write $\mathbf{U}_0 = \mathbf{U}^{(0)}$. All errors and probe responses are evaluated using the solver-state channels $(\rho, u, v, \rho E, \tilde{\nu})$ and the fixed norm $\| \mathbf{d} \|_X^2 = (5N_c)^{-1} \sum_{j=1}^{N_c} \sum_{q=1}^{5} (d_{qj}/s_q)^2$, where $s_q$ are positive, training-derived channel scales. All initializations, references, and probe states are passed through a common solver-compatible state map before evaluating the ADflow-aligned residual and Jacobian actions.

For the idealized full-step correction, with $\mathbf{e}_k = \mathbf{U}^\star - \mathbf{U}^{(k)}$ and $E_k = \| \mathbf{e}_k \|_X$, define $\mathbf{g}_k = \mathbf{J}_k^{-1}[\mathbf{R}_k + \mathbf{J}_k \mathbf{e}_k]$. Then

$$\mathbf{e}_{k+1} = \mathbf{e}_k + \mathbf{J}_k^{-1}\mathbf{R}_k = \mathbf{g}_k.$$

Hence, $E_{k+1} = \| \mathbf{g}_k \|_X$ and $E_1 = (1 - \chi_0)E_0$. If $E_0 = 0$, the initialization is already at the target; if $E_0 > 0$ and $\mathbf{g}_0 = \mathbf{0}$, the first correction reaches the target. In both cases, the subsequent error is zero, and singular diagnostic ratios need not be evaluated. The following analysis considers $E_0 > 0$ and $E_1 > 0$, with an invertible initial Jacobian and well-defined probes.

At the initial state, define the finite linearization-defect response as

$$\mathcal{D}_{\mathbf{U}_0}(\mathbf{d}) = \mathbf{J}_0^{-1}[\mathbf{R}(\mathbf{U}_0 + \mathbf{d}) - \mathbf{R}(\mathbf{U}_0) - \mathbf{J}_0\mathbf{d}].$$

Because $\mathbf{R}(\mathbf{U}^\star) = \mathbf{0}$, the initial-error probe satisfies $\mathcal{D}_{\mathbf{U}_0}(\mathbf{e}_0) = -\mathbf{g}_0$. Its quadratic coefficient and that of the generated-error probe define

$$c_0 = \frac{\| \mathbf{g}_0 \|_X}{E_0^2}, \qquad s_{g,0} = \frac{\| \mathcal{D}_{\mathbf{U}_0}(\mathbf{g}_0) \|_X}{\| \mathbf{g}_0 \|_X^2}, \qquad S_0 = \max\{c_0, s_{g,0}\}.$$

Both probes use the initial base state and inverse Jacobian, with the original displacement amplitudes. The second probe direction is the first generated error, $\mathbf{e}_1 = \mathbf{g}_0$; however, the subsequent correction uses a different base state and Jacobian. Assumption 1 relates the initial scale $S_0$ to these later responses. It applies to every generated state with $E_k > 0$, provided that the errors have remained within $E_0$, including a root different from the selected target.

More precisely, for the specified problem and initialization family with an invertible initial Jacobian and well-defined probes, Assumption 1 applies at every generated state $\mathbf{U}^{(k)}$ with $k \geq 1$, provided that $E_j \leq E_0$ for $0 \leq j \leq k$. If $E_k > 0$, the state must be admissible, have an invertible Jacobian, and satisfy $c_k = \| \mathbf{g}_k \|_X / E_k^2 \leq \beta S_0$. The shared finite coefficient $\beta \geq 1$ is fixed independently of the subsequent iterates of the query being assessed. If $E_k = 0$, the target state is retained.

To prove Proposition 1, use $S_0 \geq E_1/E_0^2$ to obtain

$$1 - \kappa_0 = \beta S_0 E_1 \geq \beta (\frac{E_1}{E_0})^2.$$

Since $\beta \geq 1$, positive $\kappa_0$ implies $E_1 < E_0$. Thus, the first iterate lies within the error range specified by Assumption 1. Suppose, inductively, that the errors have remained within this range and $E_k \leq E_1$. For $E_k > 0$, the assumption ensures that the next correction is well-defined and gives

$$E_{k+1} \leq \beta S_0 E_k^2 \leq \beta S_0 E_1 E_k = (1 - \kappa_0) E_k.$$

If $E_k = 0$, the target state is retained. Therefore, induction ensures that all subsequent errors remain within $E_1 < E_0$ and yields

$$E_K \leq E_1 (1 - \kappa_0)^{K-1} = E_0 (1 - \chi_0)(1 - \kappa_0)^{K-1} = B_K.$$

Because $0 \leq 1 - \kappa_0 < 1$, the error tends to zero. This establishes convergence to the selected target and the finite-step bound stated in Proposition 1.

**Corollary. Conditional query coverage.** Denote the bound in equation (7) by $B_K$. For a query distribution and tolerance $\varepsilon_U > 0$, let $\mathcal{A}$ denote the event that the initial diagnostics are well-defined and Assumption 1 holds. Let $\mathcal{S}_K$ denote a well-defined idealized sequence through step $K$ with $E_K \leq \varepsilon_U$. Then

$$\Pr[\mathcal{S}_K] \geq \Pr[\mathcal{A} \cap \{\kappa_0 > 0\} \cap \{B_K \leq \varepsilon_U\}].$$

**Proof.** The pointwise bound in Proposition 1 yields

$$\mathcal{A} \cap \{\kappa_0 > 0\} \cap \{B_K \leq \varepsilon_U\} \subseteq \mathcal{S}_K.$$

Taking probabilities establishes the coverage bound. Queries that are already at the target or reach it in a single correction can be included as exact successes. The fraction with $\kappa_0 > 0$ in Fig. 4 represents the initial-state criterion; finite-step accuracy coverage additionally requires $\mathcal{A}$ and $B_K \leq \varepsilon_U$.

We conducted a separate experiment to characterize the variation in the inverse-Jacobian-weighted nonlinear response along NK trajectories in the flow-field state space. The experiment included 100 geometries selected independently of the diagnostic assessment and development cohorts, with 25 geometries sampled from each geometric-distance quartile and one flow query per geometry. Along each trajectory, we measured $c_k = \| \mathbf{g}_k \|_X / E_k^2$ for full-step corrections at states satisfying $0 < E_k \leq E_0$. For each query, $C_i^{\mathrm{obs}}$ is defined as the maximum of $c_{0,i}$ and its eligible successor coefficients. The ratio $C_i^{\mathrm{obs}}/S_{0,i}$ compares the observed response scale with that captured by the initial probes. The empirical 75th percentile of these ratios gives

$$\beta = \max\{1,\ P_{75}(\frac{C_i^{\mathrm{obs}}}{S_{0,i}})\} \simeq 1.553.$$

This value characterizes the measured variation in response and remains fixed for the initialization comparisons, including those involving alternative initializations, without recalibration. The convergence statements continue to treat Assumption 1 as a condition.

Initial-state comparisons use FSB, uniform, and residual-matched pseudo-time initializations from the 200-query timing cohort. Pseudo-time fields are advanced from a uniform-flow initialization until their residuals, evaluated using the common mapped operator, reach or fall below the corresponding FSB residuals. Initial-error summaries are based on paired state errors, which are squared before aggregation.

For the damped, inexact update in equation (3), the corresponding error identity is

$$\mathbf{e}_{k+1} = (1 - \lambda_k)\mathbf{e}_k + \lambda_k \mathbf{g}_k - \lambda_k \mathbf{J}_k^{-1} \boldsymbol{\ell}_k.$$

Thus, step control and linear-solve defects contribute additional terms beyond the idealized response. The proposition and coverage corollary apply to idealized corrections, whereas Fig. 4 relates the initial-state diagnostics to the observed convergence behavior under the specified solver workflows.

## Datasets

### *Airfoil datasets*

Two-dimensional airfoils are represented using a class–shape transformation (CST) approach.[49] The offline corpus contains 99,569 flow states. The data are randomly split 90:10 by flow case into 89,612 training states and 9,957 validation states; consequently, geometries may occur in both populations. Operating conditions span approximately $\mathrm{Ma} = 0.10$–$0.85$, $\alpha = -2^\circ - 6^\circ$, and $\mathrm{Re} = 2.2 \times 10^6 - 1.9 \times 10^7$. The optimization-derived benchmark was drawn from intermediate designs generated by six-point aerodynamic optimization tasks. The benchmark follows the operating conditions of these source tasks and therefore covers a narrower range than the offline corpus: six Mach numbers ranging from 0.71 to 0.76, angles of approximately $0^\circ - 6^\circ$, and Reynolds numbers of $1.57 \times 10^7 - 1.68 \times 10^7$. The OOD assessment using this benchmark focuses on geometric variation rather than operating conditions. The representative benchmark cases in Fig. 3b and the interpolation path in Fig. 3c use $\mathrm{Ma} = 0.74$, $\alpha = 1^\circ$, and $\mathrm{Re} = 1.6378 \times 10^7$. The interpolation path connects RAE2822 to an ellipse with a thickness-to-chord ratio of 0.25.

The accuracy comparison uses 16,399 benchmark cases across 3,405 geometries. Their CFD reference solutions satisfy the residual-ratio target of $10^{-8}$ and the force-stability criteria. All returned FSB–NK endpoints are included in the error distributions, including those that terminate before convergence. Aerodynamic errors are calculated as the absolute differences from reference force coefficients. When lift and drag errors are combined, we use the weighted L1 error $E_F = |C_L - C_L^{\mathrm{ref}}| + 10|C_D - C_D^{\mathrm{ref}}|$. The combined error is calculated for each case before the population median is determined.

### *Wing datasets*

Three-dimensional wings are described by planform parameters, seven control-section shapes, and spanwise twist. The training and validation datasets are separated by geometry and contain 6,234 and 699 geometries, respectively. The paired benchmark comprises 1,085 flow cases across 624 sectional-shape-optimization geometries from the Common Research Model (CRM) and DLR-F6 section families and 318 planform-optimization geometries that vary in sweep, span, taper, or kink geometry. A 60-case subset spanning sectional-shape and planform geometries was used for paired timing. The accuracy comparison uses the original archived CFD pressure field and force coefficients for each of the 1,085 cases as the common reference for standalone FSB and the FSB–NK workflow. The three-dimensional workflow uses one ANK-to-NK solve from the FSB prediction without re-prediction. If the solve fails the residual or state checks, the original FSB prediction is used as the output for the accuracy comparison. For wings, Field MSE averages the squared error in nondimensional pressure over all volume cells. The separate surface examples use independently converged CFD solutions initialized from uniform flow.

### *Geometric distance*

For a query geometry $\mathcal{G}$, its geometric separation from the training population is measured by the mean distance to its five nearest training neighbors, normalized by the leave-one-out 99th percentile of the training distances:

$$d_5(\mathcal{G}) = \frac{1}{5} \sum_{q \in \mathcal{N}_5(\mathcal{G})} d\,(\mathcal{G}, q), \qquad d_5^*(\mathcal{G}) = \frac{d_5(\mathcal{G})}{P_{99,\mathrm{train}}}.$$

Thus, $d_5^* > 1$ denotes separation beyond the training P99 reference. For airfoils, the distance is defined as the RMS difference in chord-normalized wall coordinates. Figures 1–4 use a common reference of 27,117 supercritical training geometry records. PCA bases are fitted using training geometries only, and the displayed scores are scaled by the corresponding training-set standard deviations.

For wing sections, the distance is defined as the RMS difference in chord-normalized vertical coordinates across seven control sections, both surfaces, and common chord positions, excluding twist, chord scaling, and spanwise placement. Planform distance is defined as the Euclidean distance in the parameter space comprising sweep angle, aspect ratio, taper ratio, kink fraction, and root adjustment, with each parameter standardized by its training-set standard deviation. Both wing distance measures use the 6,234 training geometries and their respective leave-one-out P99 values. Normalized distances are interpreted within their respective geometry representations.

## Aerodynamic design optimization

### *Optimization framework*

Both tasks employ a differential-evolution population search within AeroOpt,[50] starting with 64 designs and generating 32 new candidates per generation. The transonic task runs for 50 generations and reports results from four seeds (0–3), whereas the Mars-UAV task runs for 30 generations and reports results from six seeds. All methods start from the same population for each task and seed.

Standalone FSB evaluates candidates by integrating pressure and viscous wall tractions from the predicted fields. Full NK numerically corrects every candidate using its surrogate prediction as the initial field, whereas CFD evaluates candidates using a uniform flow field as the initial state. Design performance is verified using CFD initialized from a uniform flow field, requiring residual convergence and stable force coefficients under all task conditions. The transonic CFD baseline already uses uniform-initialized CFD and therefore requires no separate re-evaluation.

***Optimization tasks***

The transonic task starts from a NACA 0012 airfoil and minimizes the mean drag across three operating conditions,

$$\min_{\mathbf{a}} C_{D,\mathrm{avg}}(\mathbf{a}) = \frac{1}{3} \sum_{m \in \{0.724, 0.734, 0.744\}} C_D\,(\mathbf{a}, m, \alpha_m),$$

with $\alpha_m$ adjusted to satisfy $C_L = 0.824$ within 0.001. The Reynolds number is determined from the reference atmospheric state. The constraints include quarter-chord pitching moment ($C_m \geq -0.092$), sectional area, thickness, leading-edge radius, angle of attack, and geometric validity. The Mars-UAV task is motivated by the airfoil-design study for Martian flight by Aprovitola et al.[37] It starts from a thin, cambered airfoil and maximizes the peak $C_L/C_D$ over nine angles of attack, ranging from $-2^\circ$ to $6^\circ$, at Ma $= 0.6$ and Re $= 20{,}000$. The constraints include sectional area and surface regularity, a maximum thickness of 0.8%–2.0% of the chord, and a maximum camber of 3.0%–6.5% of the chord.

***Top-K evaluation***

Top-$K$ evaluation couples surrogate screening with the differential-evolution search in AeroOpt. All 64 initial designs are first evaluated using FSB–NK to establish a numerically evaluated population. In each subsequent generation, AeroOpt proposes 32 new geometries from the retained designs. FSB estimates their task objectives and aerodynamic constraints, while geometric constraints are evaluated directly. These estimates are used to screen the new proposals for numerical evaluation, selecting $K = 4$ or 12 geometries. Here, $K$ denotes the number of designs, not the number of individual flow conditions: each selected transonic design is evaluated at all three Mach numbers with the required lift adjustment, whereas each selected Mars-UAV design is evaluated over the nine prescribed angles of attack.

For each selected geometry, FSB–NK corrects the predicted fields and recomputes the aerodynamic quantities used to evaluate the task objectives and constraints. The resulting numerical values replace the corresponding surrogate estimates. Unselected proposals are not added to the numerical archive and cannot compete for parent or elite membership based solely on surrogate scores. If $\mathcal{A}_g$ denotes the archive of numerically evaluated designs and $\mathcal{C}_g$ denotes the newly corrected designs, the archive update is given by

$$\mathcal{A}_{g+1} = \mathcal{A}_g \cup \mathcal{C}_g, \qquad |\mathcal{C}_g| = K.$$

AeroOpt selects the next parent population from the newly obtained numerical results and retained numerical evaluations, rather than replacing all parents with only the $K$ newly corrected designs. This retained pool maintains a 32-design parent population even when only four new designs undergo numerical correction. The elite archive records the best feasible designs based on their numerical objectives and constraint values, and the selected parents are used to generate the next set of proposals.

For an initial population of $N_0$, $G$ generations, and $N_{\text{new}}$ proposals per generation, the search schedules

$$N_{\text{corr}}^{\text{Top-}K} = N_0 + KG, \qquad N_{\text{corr}}^{\text{Full NK}} = N_0 + N_{\text{new}}G.$$

Both searches explore the same number of proposed geometries; screening reduces the number that undergo numerical correction. These counts refer to design evaluations during optimization. The selected designs are then independently verified using CFD initialized from a uniform flow field, under the same requirements as for the other methods.

**Computational cost and timing**

For the solver comparisons, CPU computations are performed using Intel Xeon Gold 6530 processors, whereas surrogate inference is performed using NVIDIA GeForce RTX 4090 GPUs with 24 GB of memory. Two-dimensional solver comparisons use eight Message Passing Interface (MPI) ranks per query, with numerical-library threading restricted to one thread per rank. Solver time includes all multigrid, ANK-to-NK, and re-prediction stages. The two-dimensional convergence comparisons use solver time. Surrogate inference and field transfer together took on the order of 0.1 s in a representative case.

Figure 4a compares FSB and residual-matched pseudo-time initializations, and NK with pure Runge–Kutta pseudo-time marching from the same FSB field. Figure 4c groups paired endpoint residual ratios according to geometric distance. The training-data study uses the same 200 queries to compare models trained on 0.1%, 1%, 10%, and 100% of the training data using the ANK-to-NK workflow, with a time limit of 120 s and a limit of 2,000 minor iterations per query. Solver work $W$ is measured as the accumulated number of minor iterations. The aggregate timing comparison uses the ratio of median solver times at termination over all 200 queries, including budget-limited runs. The threshold time is defined as the time at which the population median residual ratio first reaches $10^{-8}$.

The three-dimensional timing study performs paired solves sequentially using the same 16 physical CPU cores within each pair, with one MPI rank per core, one thread per rank, and a balanced method order. Convergence histories and timing summaries use 60 pairs in which both solves reached the target. Paired speedup is the median of their CFD-to-FSB–NK time ratios. Solver work includes cell-weighted multigrid work, as well as Krylov and residual work on the finest grid. Solver histories are aligned according to progress within each grid stage. Coarse-grid reductions are mapped continuously onto the finest-grid display scale, while preserving the recorded finest-grid residual ratios relative to the uniform-state reference for each case.

The transonic CFD, Full NK, Top-4, and Top-12 runs each used 48 CPU cores. FSB, Full NK, Top-4, and Top-12 each used one GPU for inference. Transonic optimization time covers active optimization from generation 0 through generation 50. For Mars-UAV, optimization time is estimated from active runs. Both timings exclude independent CFD verification. Optimization speedups are calculated as ratios of seed-mean times.

## Acknowledgements

This work was supported by the National Natural Science Foundation of China (grant numbers U23A2069, 12372288, U2541235 and 12388101), the National Key Research and Development Program of China (grant number 2024YFB4205601) and other national research projects.